\PassOptionsToPackage{table,dvipsnames}{xcolor}
\documentclass[sigconf]{acmart}

\AtBeginDocument{%
  }

\setcopyright{none}
\acmDOI{}
\acmISBN{}

\acmConference[LOCO '24]{1st International Workshop on Low Carbon Computing}{3rd December, 2024}{Glasgow, Scotland}

\usepackage{multirow}
\usepackage{verbatim}
\usepackage{adjustbox}
\usepackage{fancyvrb}
\usepackage{listings}
\usepackage{array}
\newcolumntype{R}[1]{>{\raggedleft\let\newline\\\arraybackslash\hspace{0pt}}m{#1}}

\newcommand{\midsepremove}{\aboverulesep=0mm \belowrulesep=0mm}
\midsepremove
\newcommand{\midsepdefault}{\aboverulesep=0mm \belowrulesep=0mm}
\midsepdefault

\newcommand{\subsub}[1]{\noindent\textbf{#1.} }

\usepackage{tabularx}

\begin{document}

\title{Ichnos: A Carbon Footprint Estimator for Scientific Workflows}

\author{Kathleen West}
\affiliation{%
  \institution{University of Glasgow}
  \city{Glasgow}
  \country{United Kingdom}
}
\email{kathleen.west@glasgow.ac.uk}

\author{Magnus Reid}
\affiliation{%
  \institution{University of Glasgow}
  \city{Glasgow}
  \country{United Kingdom}}
\email{2668261r@student.gla.ac.uk}

\author{Yehia Elkhatib}
\affiliation{%
  \institution{University of Glasgow}
  \city{Glasgow}
  \country{United Kingdom}}
\email{yehia.elkhatib@glasgow.ac.uk}

\author{Lauritz Thamsen}
\affiliation{%
  \institution{University of Glasgow}
  \city{Glasgow}
  \country{United Kingdom}}
\email{lauritz.thamsen@glasgow.ac.uk}

\renewcommand{\shortauthors}{West et al.}

\begin{abstract} 
Scientific workflows facilitate the automation of data analysis, and are used to process increasing amounts of data. Therefore, they tend to be resource-intensive and long-running, leading to significant energy consumption and carbon emissions. With ever-increasing emissions from the ICT sector, it is crucial to quantify and understand the carbon footprint of scientific workflows. However, existing tooling requires significant effort from users -- such as setting up power monitoring before executing workloads, or translating monitored metrics into the carbon footprints post-execution. 

In this paper, we introduce a system to estimate the carbon footprint of Nextflow scientific workflows that enables post-hoc estimation based on existing workflow traces, power models for computational resources utilised, and carbon intensity data aligned with the execution time. We discuss our automated power modelling approach, and compare it with commonly used estimation methodologies. Furthermore, we exemplify several potential use cases and evaluate our energy consumption estimation approach, finding its estimation error to be between 3.9--10.3\%, outperforming both baseline methodologies. 
\end{abstract}

\maketitle

\section{Introduction} 
Scientists in many fields, including genomics, materials science, and remote sensing, need to analyse increasing amounts of data \cite{muirRealCostSequencing2016b, fellowsyatesReproduciblePortableEfficient2021, schaarschmidtWorkflowEngineeringMaterials2021, berrimanMontageGridEnabled2004}. Scientific workflow systems facilitate the automation of such analyses, enabling scientists to compose pipelines out of black-box tasks with data dependencies between them. 
Because these workflows are often used to process large quantities of data, they tend to be resource-intensive and long-running, leading to significant energy consumption and, therefore, carbon emissions. Furthermore, the growing popularity of big data applications has been identified as a driver of the increasing emissions of the ICT sector~\cite{freitagRealClimateTransformative2021b}. As such, it is crucial to quantify and understand the carbon footprint of scientific workflows. 

Scientific workflow systems such as Nextflow ~\cite{ditommasoNextflowEnablesReproducible2017a} allow for the design, execution, and monitoring of workflows on heterogeneous clusters. 
While these systems usually generate detailed performance traces and logs for executed workflows, they do not produce a record of the energy consumed or carbon emitted. Consequently, users must manually monitor power consumption with hardware/software power meters or, otherwise, use a methodology like Cloud Carbon Footprint (CCF)\footnote{\label{ccf-footnote} \url{https://www.cloudcarbonfootprint.org/docs/methodology/}} or Green Algorithms (GA)~\cite{https://doi.org/10.1002/advs.202100707}, which employ linear power models to translate resource utilisation into energy consumption. In either case, to translate the energy consumed into carbon emitted, users need a measure of carbon intensity (CI), such as a yearly average or a more fine-grained metric. Generally, CI measures the amount of carbon ($CO_2e$) produced per kilowatt-hour ($kWh$) of electricity consumed, and varies across different locations, seasons, and times, depending on the sources generating electricity and the demand on the grid.

In practice, monitoring power consumption requires the user to attach a physical power meter or to enable a software-based tool like Intel's Running Average Power Limit (RAPL) prior to executing a workflow. Without this step, power consumption can only be estimated based on coarse-grained resource utilisation averages. 
This is possible using the CCF and GA methodologies, but only at reduced accuracy. 
The GA methodology relies on vendor-specified Thermal Design Power (TDP) of assigned compute resources, a proprietary metric that does not reflect key processor settings, such as processor frequency, and does not indicate idle power consumption. 
The CCF methodology builds a linear power model between the power consumption measured at 0\% and 100\%; however, this does not allow for the potential of non-linear increases in power consumption over this range~\cite{JIN2020114806}. 
Furthermore, while both methodologies translate power consumption into carbon emissions, they use a static average value to represent the CI of electricity consumed by the compute workload, ignoring the substantial temporal variability in CI.

To address these limitations, we propose \emph{Ichnos}, a novel and flexible system for estimating the carbon footprint of Nextflow workflows based on detailed workflow traces, resource-specific power models, and CI time-series data. First, Ichnos takes as input the automatically generated workflow trace produced by Nextflow. The use of these traces is an original contribution, ensuring that users do not need to manually monitor power consumption and enabling the analysis of previously executed workflows.
Next, Ichnos enables users to automatically generate a power model for utilised compute resources to accurately reflect processor settings, such as processor frequency, instead of solely relying on a linear function between min/max power consumption values per processor -- though this is offered as a fallback methodology if users no longer have access to compute resources to generate power models.  
Finally, Ichnos converts the estimated energy consumption to overall carbon emissions using fine-grained time-series CI data for each workflow task and only resorts to coarse-grained yearly averages where high-resolution location-based CI data are not available. 
Additionally, Ichnos reports estimated energy consumption and carbon emissions per workflow task, providing greater granularity than existing methodologies, and allows users to identify which of their tasks have the largest footprint to address.
We provide the implementation of Ichnos as open-source\footnote{\url{https://github.com/GlasgowC3lab/ichnos}}.
We evaluate the accuracy of the automated power models generated with Ichnos used to estimate energy consumption and compare them with the monitored energy consumption and baseline methodology estimations from CCF and GA.
We demonstrate our estimator system on traces from three real-world Nextflow workflows and show the system's functionality by varying the granularity of provided CI data, using both average and marginal CI, and varying processor governor settings of assigned compute resources.

\section{Background}
Here, we summarize scientific workflows and carbon intensity. 

\subsection{Scientific Workflows}  
Scientific workflows automate data analysis processes that support a scientific objective. They are often defined in terms of their tasks and data dependencies, and are represented using directed acyclic graphs, or as pipelines. 

\begin{figure}[h]
  \centering
  \includegraphics[width=0.9\linewidth]{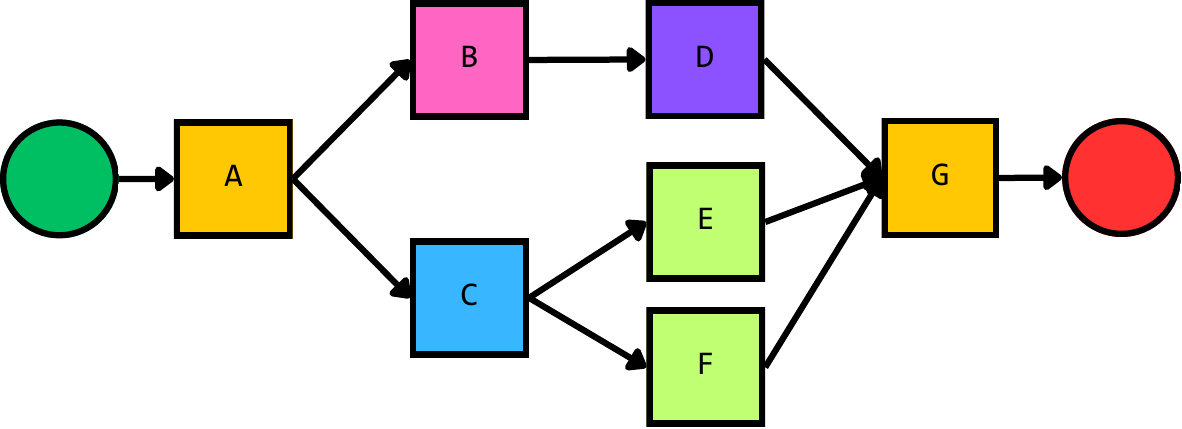}
  \caption{Simple representation of a scientific workflow, composed of seven tasks (A--G).}
  \label{fig:basic-workflow}
\end{figure} 

Figure~\ref{fig:basic-workflow} shows a simple representation of a scientific workflow, with seven tasks (A--G) and the dependencies between them. The tasks that form a workflow are considered black-box processes, receiving input from previous task(s), undergoing some processing and producing output to send on to subsequent tasks. 
As tasks are individual processes, they could run in parallel on available compute devices. For example, Tasks B and C could run in parallel after receiving input from Task A. 
Dedicated resource usage data could therefore be generated for each task, such as its runtime, status and the compute resource it was executed on. 

Scientific Workflow Management Systems (SWMS) like Kepler, Nextflow, and Pegasus allow scientists to design, execute, and monitor workflows on heterogeneous infrastructure. Such systems enable the generation of performance traces, at a task level. 

\subsection{Carbon Intensity Signals}
The carbon intensity (CI) of electricity measures carbon dioxide released per unit of energy produced. 
Generating energy from renewable energy sources such as wind or solar reduces CI. CI also typically decreases during periods of time when the demand on the grid is low. 
In this work, we use grams of carbon dioxide per kilowatt-hour ($gCO2e/kWh$) to measure CI. 

CI can be quantified using two signals: average and marginal. 
Average reflects the overall grid emissions at the time when electricity is requested, factoring in each energy source's relative share and emission rate. 
Marginal measures the emissions of the specific energy source(s) used to meet additional load at the time when electricity is requested. 
While marginal CI is preferred for measuring the impact of demand shifting, the average CI is more readily available and used for emissions accounting.
Moreover, average and marginal CI are available at different levels of granularity, such as yearly averages, hourly values, and 5/15/30-minute values. 
As a general-purpose system estimating the carbon footprint, we leave the CI signal choice to the user to provide the greatest flexibility. 

\section{System Design} 
In this section, we discuss the requirements for the design of Ichnos and also provide an overview of the estimator system's design. 

\subsection{Requirements}
We identify the following requirements from which we derive the design of Ichnos.  \\

    \subsub{Enable post-hoc estimation} We enable post-hoc use of Ichnos after workflows have been executed. This enables users to analyse the carbon footprint of previous \textbf{and} new experiments, from executions that may have occurred on local devices, clusters, and cloud infrastructure. Given this, users may no longer have access to the infrastructure used to execute  workflows, whether the devices were replaced, or only temporarily booked in a cloud environment.   \\

    \subsub{Use resource utilisation data} Resource utilisation monitoring is often more readily available than power monitoring. %
    Users typically lack access to power monitoring tools in public cloud environments. Also, users may not have configured power monitoring when workflows were executed, but will often still have access to monitoring data. Hence, Ichnos uses existing workflow traces that contain task-level resource usage data, without energy measurements.    \\

    \subsub{Estimate CPU and Memory Energy Consumption} We focus on the energy consumption of CPU and memory  as this typically has the largest dynamic power consumption range to attribute to specific load on compute resources. In addition, workflows are often executed on shared resources, where storage access may not be limited to the nodes a workflow's tasks execute on. However, workflow traces \emph{are} commonly limited to resource usage by tasks on specific nodes. \\

    \subsub{Estimate Operational Carbon Emissions} We currently estimate the operational carbon emissions. While electricity grid emissions data, such as average and marginal CI, are increasingly available, Life Cycle Assessments (LCAs) detailing embodied carbon emissions are less widely available for specific server hardware. Furthermore, for virtual resources in public cloud infrastructure, we do not have detailed information on the utilised server hardware. \\

    \subsub{Estimate Workflow Carbon Footprint} Nextflow workflow traces are produced for individual workflow runs, and offer no insight into other processes running on the same machines. We, therefore, consider the estimation of the overall system carbon footprint in multi-tenant compute scenarios, such as public cloud infrastructure, to be outside the scope of our system, and rely on the generated workflow traces for accounting for the correct shares of an overall system.  

\subsection{System Overview}
\begin{figure}[htbp]
  \centering
  \includegraphics[width=1\linewidth]{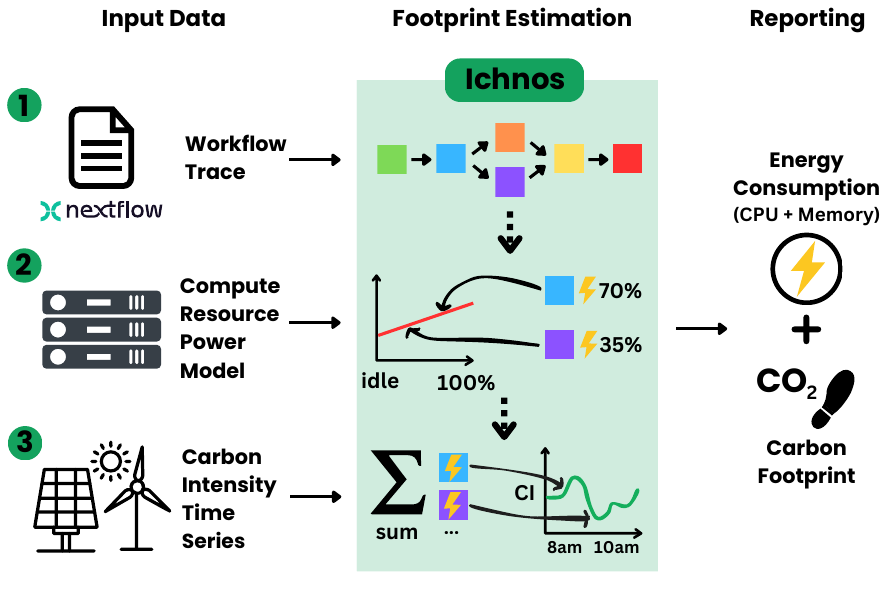}
  \caption{High-level design of the Ichnos Carbon Footprint estimator system with per-task power and emissions estimation, based on provided input data, and detailed reporting.}
  \label{fig:design}
\end{figure}

Ichnos is a system that produces an estimate of the operational carbon footprint from the execution trace of a Nextflow scientific workflow using power and energy data aligned with the execution. Figure \ref{fig:design} provides an overview of the design. 

In Phase 1, the user provides input in the form of three items:
\begin{enumerate}
    \item The workflow trace that includes a task-level summary of resource usage including the runtime, CPU utilisation, and allocated memory;
    \item The power model selected to estimate the power consumption %
    -- to accurately estimate energy consumption, the user can select an automatically generated power function or regression-based model to reflect processor settings; and
    \item The CI data which should be fine-grained time-series data, if available, or instead a coarse-grained average.
\end{enumerate}

In Phase 2, resource usage data are extracted from the workflow trace for each task, and the energy consumption is estimated using the selected power model. Subsequently, the energy consumption per task is translated into carbon emissions using the provided CI data. This estimates operational carbon emissions by aligning the tasks of potentially long-running workflow applications with CI data matching the specific execution times. These estimations are summed to calculate the power consumption and carbon emissions for the overall workflow execution. 

In Phase 3, the energy consumption and carbon emissions estimated for each task are summarised in a trace file, alongside a summary of the overall carbon footprint. We also produce a trace file identifying the 10 most energy-intensive and the 10 longest-running workflow tasks, allowing users to review their relative emissions, and to consider the potential of aligning tasks with fluctuating CI -- applying existing carbon-aware computing methods to reduce the overall footprint~\cite{wiesner2021let, hanafy2023carbonscaler}. 

\subsection{Automated Power Modelling}
\label{subsec:power-modelling-explanation}
Ichnos supports generating power models for utilised compute resources in an automated manner -- selecting the most accurate available model. 
The estimator system does this in a two-step procedure, first taking several measurements of energy consumed by CPU and memory by stress-testing the CPU and memory using the Turbostress\footnote{\url{https://github.com/teads/turbostress}\label{turbo}} tool. Ichnos has a default setting of 11 measurements, adjusting CPU load from 0\% (idle) to 100\% in 10\% increments. These measurements should ideally be taken at the time of execution, on the compute resources where workflows were executed, documenting processor settings such as the governor selected (which decides how the CPU frequency is adjusted based on CPU demand), and the date to version readings. 
The second step involves generating a power model from the readings and fitting a polynomial or linear model from these readings. The power modelling phase should be run at regular intervals through compute resource lifetimes as well as when hardware is changed, to account for altered device performance. 
We evaluate the accuracy of Ichnos's power model generation in Section~\ref{subsec:power-modelling}.

In the scenario where users estimate energy consumed by a historical workflow, executed on resources that they no longer have access to, or on public cloud resources that have been released, or anywhere that a user cannot execute the energy measurements script -- Ichnos has the fallback option of using a linear power model, or if only the CPU model is known, a per-core value based on vendor-specified TDP.
Both these fallbacks are used in existing estimation methodologies and can offer ballpark estimates of the energy consumption to then translate into carbon emissions. 
The memory energy consumption coefficient in Ichnos can be configured according to the Turbostress readings. Otherwise, the tool defaults to a constant conversion factor, as used in existing estimation methodologies. The memory stress testing is less comprehensive than that of the CPU. However, we found that memory energy consumption generally has a relatively small impact on the overall workflow energy consumption.

\subsection{Carbon Footprint Reporting}
Ichnos offers human-readable files to quickly understand the footprint of a workflow for general reporting. It also produces computer-readable files as an augmented trace file where the tasks' carbon emissions and energy consumption are reported. This is provided to enable scientists to better understand their workflow's footprint, and the heavy-hitting tasks that may be disproportionately contributing to the overall footprint. 

The following three files are generated: 
\begin{enumerate}
    \item a human-readable summary of the workflow, detailing parameters provided and overall energy consumption and carbon emissions;
    \item a computer-readable file with the carbon emissions and energy consumption estimations of individual tasks;
    \item a computer-readable summary with the top 10 tasks that had the highest carbon footprint and energy consumption.
\end{enumerate}

Table~\ref{table:task-ranked} shows an extract from the summary listing the top 10 tasks with the highest carbon footprints and energy consumption. 
In the example, we can identify tasks that contribute to the overall footprint, with, for example, PICRUST consuming 0.034$kWh$, almost 6x as much as BARRNAP, and, since both are executed on similar carbon-intensive energy, a similarly larger carbon footprint.

\begin{table}[t]
\centering
\resizebox{\columnwidth}{!}{
\begin{tabular}{llrrrlrrr}
\toprule
name              & id & co2e  & energy & avg\_ci        & realtime  & cores & usage (\%) \\ %
\midrule
DADA2\_ERR        & 58 & 2.742 & 0.044  & 55.0           & 2576181    & 6          & 306.0      \\ %
PICRUST           & 66 & 2.337 & 0.034  & 60.5           & 1977266    & 6          & 340.0      \\ %
DADA2\_DENOISING  & 59 & 0.678 & 0.011  & 56.0           & 651643     & 6          & 276.0      \\ %
BARRNAP           & 65 & 0.347 & 0.006  & 56.0           & 348293     & 2          & 198.0      \\ %
DADA2\_ADDSPECIES & 67 & 0.285 & 0.005  & 56.0           & 288655     & 1          & 100.0      \\ %
DADA2\_TAXONOMY   & 64 & 0.278 & 0.005  & 56.0           & 204224     & 16         & 1310.0     \\ %
DADA2\_RMCHIMERA  & 60 & 0.074 & 0.001  & 56.0           & 66396      & 6          & 506.0      \\ %
DADA2\_QUALITY1   & 40 & 0.054 & 0.001  & 49.0           & 63627      & 2          & 102.0      \\ %
DADA2\_QUALITY1   & 39 & 0.054 & 0.001  & 49.0           & 63645      & 2          & 101.0       \\ %
MULTIQC           & 70 & 0.052 & 0.001  & 65.0           & 48646      & 1          & 57.0        \\ %
\bottomrule
\end{tabular}
}
\caption{Extract from detailed summary file listing tasks with the highest carbon footprint and energy consumption.}
\label{table:task-ranked}
\end{table}

\section{Evaluation}
\label{sec:evaluation}
In this section, we first present the experimental setup that we used to evaluate Ichnos. We then analyse the accuracy of the estimator system's generated power models for estimating power consumed during workflow execution, and demonstrate a series of practical use cases for Ichnos.

\subsection{Experimental Setup}
We use a set of baselines, workflows, and infrastructure for multiple experiments as described here.

\paragraph{Baselines}
To evaluate the accuracy of our approach when estimating energy consumption -- and therefore carbon emissions --, we compare it against two baseline methodologies that also enable post-hoc footprint estimation: Naive--Linear and Green Algorithms (GA). 
Naive--Linear is aligned with the Cloud Carbon Footprint (CCF) estimation methodology\textsuperscript{\ref{ccf-footnote}}, assuming that energy consumption scales linearly between the idle and maximum power consumption reported by manufacturers or online databases like SPEC\footnote{\url{https://www.spec.org/power_ssj2008/}}. 
GA uses the manufacturer-specified processor TDP to estimate per-core energy consumption. 
In addition to comparing against both baselines, we compare all estimations against power consumption readings taken with the Perf tool, which uses Intel's RAPL, which we consider to be the ground truth in our experiments. 

\paragraph{Workflows}
To evaluate our system, we selected two real-world workflows from the nf-core repository\footnote{\url{https://github.com/nf-core}} -- a community-curated collection of workflows implemented using Nextflow~\cite{ewels2020nf}. We chose the AmpliSeq and NanoSeq workflows, both are bioinformatics workflows within the top 10 most popular workflows from nf-core. We manually executed these workflows to produce experimental data with energy consumption monitoring. In addition, we exemplify the system's capacity for post-hoc carbon footprint estimation by using historical traces where the Rangeland workflow was executed on an infrastructure we do not have access to. 

\paragraph{Infrastructure}
For the following power modelling accuracy analysis and two of the use cases, we used five nodes from a local heterogeneous cluster -- these nodes each have the prefix \textit{gpgnode}. 
Within the cluster, gpgnodes 13--16 are each equipped with two Intel Xeon E5-2640 v2 processors, which have a 1.2--2.5$GHz$ frequency range, and 64 GB of RAM. 
In contrast, gpgnode 22 is equipped with an Intel Xeon Gold 6426Y processor, which has a frequency range of 0.8--4.1$GHz$, and 128 GB of RAM.
Meanwhile, we describe the infrastructures relevant for the other use case demonstrations in the respective subsections.

\subsection{Power Modelling Accuracy}
\label{subsec:power-modelling}
We discuss the accuracy of Ichnos's generated power models, in relation to the actual energy consumption, before comparing our estimations with the selected baselines of Naive--Linear and GA. 

\paragraph{Power Model Accuracy}
For each available compute device, we took power consumption readings, as detailed in Section~\ref{subsec:power-modelling-explanation}. We used these readings to generate cubic and linear regression models of consumed energy.

\begin{figure*}[t]
  \centering
  \includegraphics[width=\linewidth]{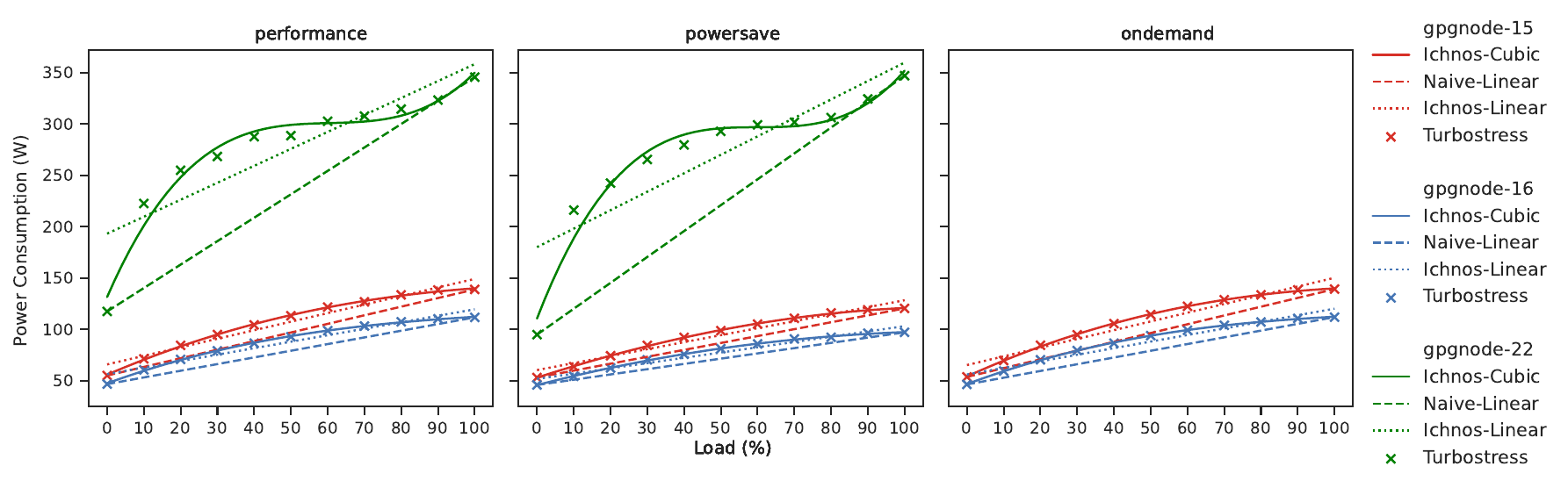}
  \caption{Comparison of power consumption readings over relative load for three compute nodes utilising different Intel governors (as available on processors).}
  \label{fig:graph-nodes-compare-models}
\end{figure*}

In Figure~\ref{fig:graph-nodes-compare-models}, we show a plot comparing power models generated for three compute nodes (gpgnode 15, 16, and 22), using the governors: performance, powersave, and ondemand -- where available. 
The original Turbostress\textsuperscript{\ref{turbo}} readings are marked on the plots. To these readings, Ichnos fits a cubic model (Ichnos--Cubic) and a linear one (Ichnos--Linear). We additionally plot the Naive--Linear model, which assumes linear scaling of power consumption from the readings at 0\% (idle) and 100\% (peak) utilisation.  

Despite these nodes being equipped with the same resources, their peak power consumption can vary substantially. For example, gpgnode-15 using the performance governor reached \textasciitilde140$W$ at 100\% utilisation while gpgnode-16 reported a value \textasciitilde110$W$ at the same utilisation, using the same governor settings. This difference ($\approx$27\%) underscores the importance of resource-specific power modelling based on actual measurements. 
The selected governor also impacts the power model, with the powersave governor consistently using less power at 100\% utilisation. 

To further demonstrate how hardware generation affects power consumption patterns, readings for gpgnode-22 are plotted. This newer hardware not only consumes significantly more power as load is increased from 0--100\%, but also exhibits markedly different behaviour.
We observe the non-linear nature of the power consumed in this node in particular, which highlights why linear models can be inadequate.
This non-linearity is especially pronounced in the 60--80\% utilisation range for this node, where the power consumption does not increase significantly with utilisation. 

We also report the Root Mean Square Error (RMSE) between the model predicted values and the energy consumption readings taken using Turbostress (which uses RAPL to take measurements), for Ichnos--Cubic, Ichnos--Linear and Naive--Linear models. These are listed in Table~\ref{table:models-energy-error}. 
Across all selected nodes, Ichnos--Cubic demonstrated superior accuracy, markedly outperforming Ichnos--Linear which, in turn, proved more accurate than the Naive--Linear model. 

\paragraph{Use of Generated Power Models to Estimate Energy Consumption}
We executed the Ampliseq workflow on gpgnodes 13--16 and 22, on selected governors and monitored the energy consumed. We used Ichnos to generate estimations from the workflow traces using the Ichnos--Cubic, Ichnos--Linear, Naive--Linear, and GA models. 

For each node using each specified processor governor setting, we executed the workflow 3 times and report the mean energy consumption. We compare the estimated energy consumption, with the actual consumption recorded using Perf, and report the overall percentage error in Table~\ref{table:energy-estimation-w-error}. We found that Ichnos--Linear performed the best for all our estimations, in line with our expectations. 

\begin{table}[tb]
\caption{The RMSE of generated models energy consumption estimations compared to Turbostress\textsuperscript{\ref{turbo}} readings.} 
\label{table:models-energy-error}
\rowcolors{3}{}{Green!30}
\centering
\begin{tabular}{clrrr}
\toprule
\multirow{2}{*}{Node}       & \multirow{2}{*}{Governor}    & \multirow{2}{*}{Naive Linear} & \multicolumn{2}{c}{Ichnos} \\
\cline{4-5}
&& & Linear & Cubic \\
\midrule
gpgnode-13 & ondemand    & 12.96  & 6.35          & 0.59         \\
" & performance & 12.14  & 5.83          & 0.84         \\
" & powersave   & 7.95   & 4.02          & 0.22         \\
\midrule
gpgnode-14 & ondemand    & 12.21  & 5.97          & 0.61         \\
" & performance & 11.55  & 5.49          & 0.86         \\
" & powersave   & 7.45   & 3.77          & 0.29         \\
\midrule
gpgnode-15 & ondemand    & 12.96  & 6.36          & 0.62         \\
" & performance & 12.01  & 5.72          & 0.85         \\
" & powersave   & 8.82   & 4.43          & 0.48         \\
\midrule
gpgnode-16 & ondemand    & 10.33  & 5.13          & 0.34         \\
" & performance & 10.10  & 4.88          & 0.53         \\
" & powersave   & 6.94   & 3.52          & 0.47         \\
\midrule
gpgnode-22 & performance & 56.45  & 28.78         & 9.61         \\
" & powersave   & 63.10  & 32.34         & 10.60        \\
\bottomrule
\end{tabular}
\end{table}

\begin{table*}[]
\caption{Estimated energy consumption for Ampliseq using generated models compared to Perf readings (ground truth).} 
\label{table:energy-estimation-w-error}
\rowcolors{3}{Green!30}{}
\centering
\begin{tabular}{clr|rr|rr|rr|rr}
\toprule
\multirow{2}{*}{Node} & Governor & Perf & Ichnos--Cubic & Error & Ichnos--Linear & Error & Naive--Linear & Error & GA & Error \\
& & (kWh) & (kWh) & (\%) & (kWh) & (\%) & (kWh) & (\%) & (kWh) & (\%) \\
\midrule
gpgnode-13 & ondemand    & 0.161 & 0.135 & 16.1 & 0.144 & 10.3 & 0.121 & 24.7 & 0.28  & 82.9 \\
\midrule
gpgnode-14 & performance & 0.161 & 0.138 & 14.2 & 0.146 & 9.1  & 0.124 & 22.8 & 0.026 & 83.7 \\
" & powersave   & 0.159 & 0.143 & 9.8  & 0.150 & 5.6  & 0.136 & 14.4 & 0.029 & 81.4 \\
\midrule
gpgnode-15 & performance & 0.168 & 0.147 & 12.4 & 0.155 & 7.4  & 0.134 & 19.9 & 0.027 & 83.6 \\
" & powersave   & 0.178 & 0.157 & 11.7 & 0.165 & 7.3  & 0.148 & 16.7 & 0.031 & 82.4 \\
\midrule
gpgnode-16 & ondemand    & 0.139 & 0.124 & 10.8 & 0.131 & 5.4  & 0.113 & 18.8 & 0.026 & 81.4 \\
\midrule
gpgnode-22 & performance & 0.165 & 0.131 & 20.7 & 0.159 & 3.9  & 0.101 & 38.7 & 0.003 & 98.0 \\
" & powersave   & 0.163 & 0.031 & 81.0 & 0.150 & 8.0  & 0.085 & 47.9 & 0.003 & 98.0 \\
\bottomrule
\end{tabular}
\end{table*}

\paragraph{Discussion of the Experimental Accuracy of Energy Estimation}
In our evaluation of the accuracy of generated power models, Ichnos--Cubic was found to be the most accurate -- reporting the minimum RMSE between model predictions and Turbostress readings. However, when we used the power models to estimate energy consumption for real Nextflow executions and reported the percentage error between the estimated and actual energy consumption, we found that Ichnos--Linear consistently outperformed Ichnos--Cubic. 

Given these results, we identify the following two drawbacks of using a non-linear model like Ichnos--Cubic to estimate energy consumption. 

\begin{figure}[bt]
  \centering
  \includegraphics[width=\linewidth]{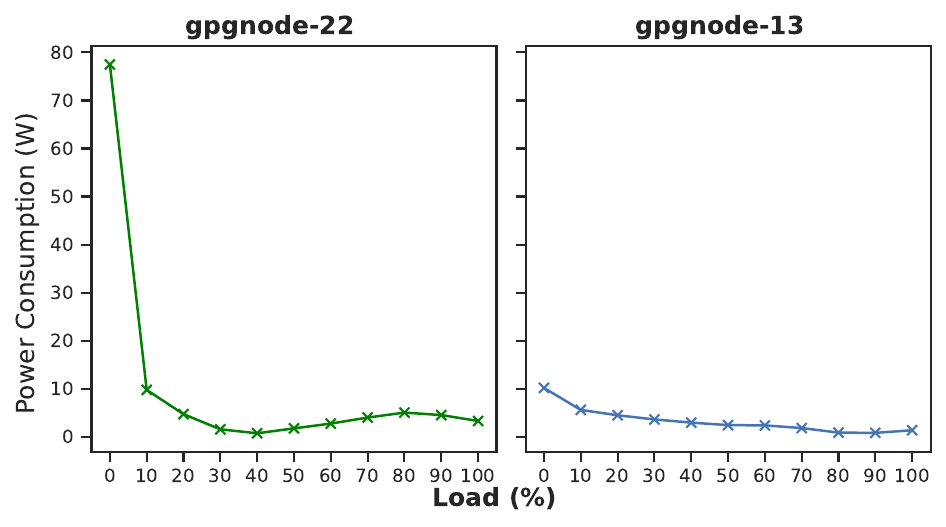}
  \caption{Comparison over two nodes of power consumed by a task as background system load varies.}
  \label{fig:background-task-impact}
\end{figure}

To enable trace-based resource estimations, we estimate the energy consumption for all individual tasks, summing each task's share of utilised resources. This works when we use a linear model for several tasks running in parallel on the same shared compute resource. 
However, if we use a cubic model and consider a task that has an average CPU utilisation of 100\% on one core, its power consumption will differ depending on whether it runs on its own or shares resources with other CPU-intensive tasks. This effect can be seen in Figure~\ref{fig:background-task-impact}, which shows the power consumption of the commonly used bioinformatics task FastQC\footnote{\url{https://github.com/s-andrews/FastQC}} at different system background loads. This task fully utilises one core on each system. We can see that the power consumption is significantly affected by the background system load, and this is especially notable for gpgnode 22, which shows a large difference between 0--10\% utilisation. This result is expected, and reflects the behaviour seen in Figure~\ref{fig:graph-nodes-compare-models}.

Furthermore, the traces generated from Nextflow executions only provide coarse-grained CPU utilisation averages for each individual workflow task -- which could have a runtime spanning seconds to hours in length and markedly different utilisation over time. 

Therefore, we recommend use of the generated Ichnos--Linear model when estimating energy consumption to reduce the power modelling estimation error -- compared to Naive--Linear and GA -- and avoid the estimation being affected by background task utilisation and how resource utilisation data are aggregated. 

\subsection{Ichnos Use Cases}  

\subsub{Use of Historical Traces}  
We used Ichnos to estimate the carbon footprint of historical executions reported for the Rangeland workflow. Specifically, we estimated the carbon footprint of the three workflow executions that occurred on a single node in Germany equipped with an Intel Xeon Silver 4314 processor and 256 GB of RAM. We configured the system to use the yearly average CI in Germany in 2023 as 394$gCO_2e/kWh$. As we did not have access to perform power measurements on the node at the time of execution, we used a linear power model ranging between 80--135W to estimate energy consumption. The average energy consumption was \textbf{30.51$kWh$}, with CPU energy consumption accounting for 99\% of the overall energy consumption and memory responsible for the remaining 1\%. The estimated carbon footprint was \textbf{12$kgCO_2e$}.  \\

\subsub{Use of Cloud Traces}
We executed the NanoSeq workflow using the Google Cloud Batch execution environment with Nextflow. The workflow ran on n2 and c2 cloud nodes in the europe-west-2 region. We had access to the workflow trace generated, but did not have access to the utilised nodes to retrieve energy measurements to fit a power model. Such a scenario may resemble scientists’ real-world use cases for Ichnos, where they have limited information on shared nodes and also on the specific location of these to access accurate CI data.
Despite these challenges, Ichnos allows for post-hoc estimation.  

We configured the system to use a linear power model, and made use of CCF's coefficients repository\footnote{\url{https://github.com/cloud-carbon-footprint/ccf-coefficients}} to obtain measurements for the Cascade Lake machine family -- used by the n2 and c2 machine types. These coefficients stated a minimum of 0.69$W$ per-core, and a maximum of 3.75$W$ per-core. We adjusted these values in line with the number of cores requested for each task in our estimations. 
We used the quoted average CI value for the europe-west-2 region that Google puts\footnote{\url{https://cloud.google.com/sustainability/region-carbon}} at 136$gCO2e/kWh$.
Using Ichnos, we estimated the carbon emissions to be 15.5$gCO2e$, translated from an estimated energy consumption of 0.114$kWh$. \\

\begin{figure}[tbp]
  \centering
  \includegraphics[width=0.9\linewidth]{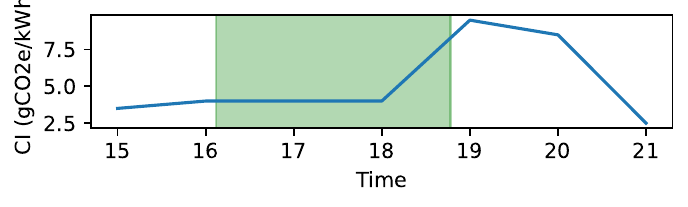}
  \caption{Variable Carbon Intensity in South Scotland on 26/09/2024, with workflow execution highlighted in green.}
  \label{fig:ci}
\end{figure}

\subsub{Varying the Granularity of CI Data}
We took the workflow trace from the AmpliSeq execution that ran on the edge server, equipped with an Intel i7-10700T processor with 32 GB of RAM, for 2h 40m in the evening of September 26\textsuperscript{th} 2024 in Glasgow. The CI fluctuation for South Scotland region of the National Grid\footnote{\url{https://carbonintensity.org.uk/}} is depicted in Figure \ref{fig:ci}. If this information was not available, we would use the average CI for the National Grid in 2023, which was 215$gCO2e$\footnote{\url{https://app.electricitymaps.com/}}. 

Ichnos estimated that the footprint was \textbf{0.33$gCO2e$} using the region-specific time-series CI data, whereas the footprint estimated using the coarse-grained average would be \textbf{18.65$gCO2e$} - an estimate almost 60x larger. 
This highlights the potential of using a flexible system, such as Ichnos, where the user can provide specific high-resolution time-series CI data to estimate the carbon footprint more accurately. \\

\subsub{Varying Processor Governor Settings}
We executed the AmpliSeq workflow using the full-size real-world dataset provided from nf-core on gpgnodes 13--16 and 22. Each compute node used 2--3 available governors, considering the real-world scenario where users may not be in control of the governor in use, and cannot set a node to a fixed frequency. We monitored the energy consumption using Perf again for comparison. We estimated the carbon emissions using Ichnos configured with the fitted linear model (Ichnos--Linear) as discussed in Section~\ref{subsec:power-modelling}. We compare the estimations with the Naive--Linear and GA baselines. 

Table~\ref{table:carbon-emissions-by-node} lists the experimental results. We cannot directly evaluate the accuracy of our carbon footprint estimations, as Perf only monitors energy consumption, rather than carbon emissions. However, the accuracy of the selected power models can be considered. We found that energy consumption estimations from Ichnos--Linear had an error of 3.9--10.3\%, compared to Naive--Linear's error of 14.4--47.9\%, and GA's error of 81.4--98.0\%. Ichnos translated the estimated energy consumption for each of these power models using the same CI data, at the same level of granularity, carrying any estimation error linearly to the carbon footprint estimates. Therefore, Ichnos enables post-hoc estimation with a reduced estimation error, outperforming both baselines.   \\

\begin{table*}[]
\caption{The estimated energy consumption and carbon emissions using Ichnos--Linear and baseline models.}
\label{table:carbon-emissions-by-node}
\rowcolors{3}{Green!30}{}
\centering
\begin{tabular}{clr|rr|rr|rr}
\toprule
\multirow{2}{*}{Node} & Governor & Perf  & Ichnos--Linear & Emissions & Naive--Linear & Emissions & GA & Emissions \\
& & (kWh) & (kWh) & (gCO2e) & (kWh) & (gCO2e) & (kWh) & (gCO2e) \\
\midrule
gpgnode-13 & ondemand    & 0.161 & 0.144 & 2.93 & 0.121 & 2.46 & 0.28 & 0.56 \\
\midrule
gpgnode-14 & performance & 0.161 & 0.146 & 2.96 & 0.124 & 2.51 & 0.026 & 0.53 \\
" & powersave   & 0.159 & 0.150 & 1.78 & 0.136 & 1.61 & 0.029 & 0.36 \\
\midrule
gpgnode-15 & performance & 0.168 & 0.155 & 1.79 & 0.134 & 1.54 & 0.027 & 0.31 \\
" & powersave   & 0.178 & 0.165 & 4.05 & 0.148 & 3.64 & 0.031 & 0.76 \\
\midrule
gpgnode-16 & ondemand    & 0.139 & 0.131 & 7.89 & 0.113 & 6.77 & 0.026 & 1.55 \\
\midrule
gpgnode-22 & performance & 0.165 & 0.159 & 6.52 & 0.101 & 4.16 & 0.003 & 0.13 \\
" & powersave   & 0.163 & 0.150 & 8.35 & 0.085 & 4.73 & 0.003 & 0.18 \\
\bottomrule
\end{tabular}
\end{table*} 

\subsub{Use of Average and Marginal CI Data}
Ichnos offers its users the opportunity to provide CI data in an input file of dynamic intervals of granularity. This also allows users to provide average or marginal CI data, which are often available at different levels of granularity, from different providers. 

To demonstrate Ichnos's flexibility, we show the footprint estimations for executions of the Ampliseq workflow on gpgnode-13--16 and 22, where the available governors were used. These executions occurred in February 2025, and we retrieved CI data aligning with the actual execution times. Average CI data were available for 30-minute intervals from the National Grid for the South Scotland region where the servers were located. Marginal CI data was available for 5-minute intervals from WattTime\footnote{\url{https://watttime.org/}} for the United Kingdom, yet more regional data was not available.

The estimated carbon emissions produced by Ichnos -- using the Ichnos--Linear power model -- are shown in Table~\ref{table:avg-and-marg-data}, demonstrating that the system enables footprint estimation with the most granular CI data available, offering users the choice of signal. 

\begin{table}
\caption{The estimated energy consumption and carbon emissions using average and marginal CI data.}
\label{table:avg-and-marg-data}
\rowcolors{3}{}{Green!30}
\centering
\begin{tabular}{clrrrr}
\toprule 
\multirow{2}{*}{Node} & Governor & Energy & \multicolumn{2}{c}{Emissions (gCO2e)} \\
\cline{4-5}
& & (kWh) & Average & Marginal \\ 
\midrule 
gpgnode-13 & ondemand & 0.144 & 2.93 & 141.95  \\
\midrule
gpgnode-14 & performance & 0.146 & 2.96 & 142.94  \\
" & powersave & 0.150 & 1.78 & 154.53  \\
\midrule
gpgnode-15 & performance & 0.155 & 1.79 & 159.30  \\
" & powersave & 0.165 & 4.05 & 157.93  \\
\midrule
gpgnode-16 & ondemand & 0.131 & 7.89 & 138.75  \\
\midrule
gpgnode-22 & performance & 0.159 & 6.52 & 149.92  \\
" & powersave & 0.150 & 8.35 & 156.08  \\
\bottomrule 
\end{tabular}
\end{table} 

\section{Limitations} 

\subsub{Generality of Ichnos}
Ichnos was created to estimate the carbon footprint of Nextflow workflow executions from automatically generated trace files. The system is configured with the expected format of this trace file, constructing records for each individual task. Therefore, it is not directly applicable to other workflow systems. However, the method used to estimate energy consumption, and to translate such consumption into carbon emissions using CI data could be adapted to work with performance trace file formats used for other workflow systems, like Pegasus or Makeflow. We welcome contributions to the open-source estimator system, which remains under active development.   \\

\subsub{Power Model Generation and Use}
Ichnos allows users to run a series of energy consumption readings by varying the computational load on a CPU, generating a resource-specific power model from these readings for use when estimating the carbon footprint. This power model generation process should be repeated when the hardware is changed or at regular intervals to track device degradation. However, this is reliant on the user having the correct permissions available and access to compute infrastructure on which workflows are executed. If power models are not up-to-date or do not align with the original workflow execution times, the accuracy of the energy consumption estimations will be limited.   \\

\subsub{Limitations of Input Data}
Ichnos allows users to provide CI data at varied levels of granularity -- enabling the use of both average and marginal CI. However, CI data usually specify a value over a given period of time, e.g. WattTime offers marginal CI at intervals of five-minutes, while the National Grid offers average CI at intervals of thirty-minutes. As these intervals become less granular, the overall footprint estimation becomes less accurate. Furthermore, we are reliant on these CI data sources for supplying accurate data. \\

\noindent By enabling post-hoc estimation, where the user will likely not have had access to power or emissions monitoring tools, we can only guarantee that our tool uses the estimation methodology described and the user-provided data -- the workflow trace and the CI as well as the power model generated from measurements on compute resources utilised at the time, or as close as possible. This limitation is the same for other existing footprint estimation methodologies like CCF and GA. 

\section{Related Work} 

This section examines energy consumption measurement methods in a broad context, followed by carbon footprint estimation methodologies and prior research wherein the carbon footprint of scientific workflows has been explored. 

To estimate the carbon footprint of computation, power consumption must first be monitored with power meters, or measured with energy profiling tools. These traditionally rely on software interfaces like the Running Average Power Limit (RAPL) -- available on Intel Processors -- or the NVIDIA Management Library (NVML) -- available on NVIDIA GPUs. Tools built using these interfaces, such as Nvidia-smi, Perf and Scaphandre, can provide accurate measurements of energy consumption~\cite{jay2023experimental}. However, these necessitate configuration prior to workload execution. Given the requirement that our estimations are made post-hoc, these are not suitable in our problem space. We instead focus on methods that are capable of modelling power consumption based on compute resource usage. 

Many methods have been proposed to model the power consumption at server and data-centre levels~\cite{9599719}. Some works consider the power consumption of a server to be the sum of idle consumption -- thought to be a fixed value -- and active consumption caused by computational workloads ~\cite{roy2013energy, dhiman2010system, xiao2013virtual}. Other studies use regression models to predict power consumption based on server properties in combination with the idle power consumption~\cite{10.1145/1250662.1250665, economou2006full, li2012online, zhang2013high}.  
Some works consider the CPU utilisation to be the dominant contributor when modelling server consumption (for example, the linear model given by Fan et al.~\cite{10.1145/1250662.1250665}) which produced reasonably accurate estimations and has since been implemented in various estimation methodologies, including CCF. From the many models proposed, we focused on implementing a fitted linear-regression model, a fitted cubic-regression model and compared these with models that are commonly used by existing estimation methodologies. 

Several tools have been created to estimate the carbon footprint from computational workloads~\cite{henderson2020systematic, benoit_courty_2024_11171501, eco2AI, anthony2020carbontracker, cumulator}. Many of these tools model power consumption by using the server utilisation and the TDP reported by the manufacturer~\cite{https://doi.org/10.1002/advs.202100707, benoit_courty_2024_11171501, eco2AI}. However, this value does not consider processor settings such as the frequency, reducing the estimation's accuracy. Other tools require the user to have privileged (root) access, and to configure tools prior to workload execution~\cite{henderson2020systematic, anthony2020carbontracker, benoit_courty_2024_11171501, cumulator}. With our requirement of enabling post-hoc estimation, these tools are out of scope. We focus on existing carbon footprint estimation methodologies, which also enable post-hoc estimation and involve an intermediate step where energy consumption is estimated: Cloud Carbon Footprint\textsuperscript{\ref{ccf-footnote}} (CCF) and Green Algorithms~\cite{https://doi.org/10.1002/advs.202100707} (GA). We compare Ichnos against both methodologies in our evaluation. 

Some research studies have specifically applied existing carbon footprint estimation methodologies to analyse the footprint of bioinformatics~\cite{10.1093/molbev/msac034} and neuroimaging~\cite{souter2024measuring} research processes. 
In other works where the focus is instead on reducing the energy footprint of scientific workflows, linear power models have been employed to estimate power consumption~\cite{saadiReducingEnergyFootprint2023}. 
The presence of these works clearly indicates interest in being able to estimate the carbon footprint of computation, and validate the use of estimation methodologies in a post-hoc manner.

\section{Conclusion} 
In this paper, we presented Ichnos -- a new system to estimate the carbon footprint of Nextflow workflow executions based on a given trace file, generated power models, and configured carbon intensity data. The system enables the post-hoc estimation of energy and carbon footprints. 
To improve the accuracy of the estimated power consumption, Ichnos allows for a series of power measurements to be taken to create a power model for estimating the energy consumption. These measurements are repeatable, enabling users to update the models when processor settings change or device performance degrades. 

We evaluated the accuracy of the generated power models in comparison to existing carbon footprint estimation methodologies. We showed that Ichnos reduced the estimation error to between 3.9--10.3\%, compared to Naive--Linear's 14.4--47.9\% and GA's error of 81.4--98.0\%. 
We further exemplified the estimator system's use cases, where users could estimate the footprint from historical workflow executions, or provide as granular CI data as they have access to (ranging, for example, from a yearly to five-minute resolution), enabling users to consider both average and marginal CI data. 

In the future, we plan to test our method for workflow executions on distributed, heterogeneous clusters with node-specific, versioned power models, as we currently use a single power model for workflows executed across multiple nodes. Furthermore, we will add support for the estimation of embodied carbon emissions and are keen to explore integrating Ichnos with methods that predict workflow performance (e.g.~\cite{BADER2024171}).

\section*{Acknowledgments}

This work was supported by the Engineering and Physical Sciences Research Council under grant number UKRI154. 
We also gratefully acknowledge the sources of electricity grid data: NESO Open Data and Electricity Maps historical data for average carbon intensity as well as marginal operating emission rates calculated by WattTime.

\section*{Rights Retention}

For the purpose of open access, we have applied a Creative Commons Attribution (CC BY) licence to this manuscript.

\section*{Data/Code Availability}

An open-source implementation of Ichnos is available at {\url{https://github.com/GlasgowC3lab/ichnos}}.

\newpage

\balance
\bibliographystyle{ACM-Reference-Format}
\bibliography{references}

\end{document}